\documentclass{llncs}

\usepackage{latexsym}
\usepackage{psfig}
\usepackage{url}

\newcommand{\cmpl}{\mbox{{\it complete}}}

\newcommand{\pos}{\mbox{{\it pos}}}
\newcommand{\perm}{\mbox{{\it perm}}}
\newcommand{\nn}{\mbox{{\it neg}}}
\newcommand{\stb}{\mbox{{\it stable}}}
\newcommand{\stba}{\mbox{{\it stable\_aux}}}

\newcommand{\At}{\mbox{{\it At}}}
\newcommand{\Lit}{\mbox{{\it Lit}}}

\newcommand{\n}{\mbox{{\bf not}}}

\newcommand{\lm}{\mbox{{\it lm}}}

\newcommand{\lar}{\leftarrow}

\title{Computing stable models: worst-case performance estimates}

\titlerunning{Computing stable models}

\author{Zbigniew Lonc\inst{1} \and Miros\l aw Truszczy\'nski\inst{2}}
\authorrunning{Zbigniew Lonc and Miros\l aw Truszczy\'nski}
\institute{Faculty of Mathematics and Information Science, Warsaw
University of Technology, 00-661 Warsaw, Poland\\
\and
Department of Computer Science, University
of Kentucky, Lexington,\\
KY 40506-0046, USA}

\begin{document}

\maketitle

\begin{abstract}
We study algorithms for computing stable models of propositional
logic programs and derive estimates on their worst-case performance 
that are asymptotically better than the trivial bound of $O(m 2^n)$,
where $m$ is the size of an input program and $n$ is the number of its
atoms. For instance, for programs, whose clauses consist of at most two
literals (counting the head) we design an algorithm to compute stable
models that works in time $O(m\times 1.44225^n)$. We present similar
results for several broader classes of programs, as well.
\end{abstract}

\section{Introduction}
\label{intro}

The stable-model semantics was introduced by Gelfond and Lifschitz
\cite{gl88} to provide an interpretation for the negation operator
in logic programming. In this paper, we study algorithms to compute 
stable models of propositional logic programs. Our goal is to design 
algorithms for which one can derive non-trivial worst-case performance 
bounds.

Computing stable models is important. It allows us to use logic 
programming with the stable-model semantics as a {\em computational} 
knowledge representation tool and as a declarative programming system. 
In most cases, when designing algorithms for computing stable models
we restrict the syntax to that of DATALOG {\em with negation}
(DATALOG$^\neg$), by eliminating function symbols from the 
language. When function symbols are allowed, models can be infinite 
and highly complex, and the general problem of existence of a stable 
model of a finite logic program is not even semi-decidable \cite{mnr94}. 
However, when function symbols are not used, stable models are 
guaranteed to be finite and can be computed.

To compute stable models of {\em finite} DATALOG$^\neg$ programs we 
usually proceed in two steps. In the first step, we {\em ground} an 
input program $P$ and produce a {\em finite} propositional program 
with the same stable models as $P$ (finiteness of the resulting {\em
ground} program is ensured by finiteness of $P$ and absence of 
function symbols). In the second step, we compute stable models of
the ground program by applying search. This general approach is used 
in {\em smodels} \cite{ns00} and {\em dlv} \cite{eflp00}, 
two most advanced systems to process DATALOG$^\neg$
programs. 

It is this second step, computing stable models of 
propositional logic programs (in particular, programs obtained by
grounding DATALOG$^{\neg}$ programs), that is of interest 
to us in the present paper. Stable models of a propositional logic 
program $P$ can be 
computed by a trivial brute-force algorithm that generates all subsets 
of the set of atoms of $P$ and, for each of these subsets, checks the 
stability condition. This algorithm can be implemented to run in time 
$O(m 2^n)$, where $m$ is the size of $P$ and $n$ is the number 
of atoms in $P$ (we will use $m$ and $n$ in this meaning 
throughout the paper). The algorithms used in {\em smodels} and {\em
dlv} refine this brute-force algorithm by employing effective 
search-space pruning techniques. Experiments show that their 
performance is much better than that of the brute-force 
algorithm. However, at present, no non-trivial upper bound on their 
worst-case running time is known. In fact, {\em no} algorithms for 
computing stable models are known whose worst-case performance
is {\em provably} better than that of the brute-force algorithm.
Our main goal is to design such algorithms.

To this end, we propose a general template for an algorithm to compute
stable models of propositional programs. 
The template involves an 
auxiliary procedure whose particular instantiation determines the 
specific algorithm and its running time. We propose concrete 
implementations of this procedure and show that the resulting
algorithms for computing stable models are asymptotically better than 
the straightforward algorithm described above. The performance analysis
of our algorithms is closely related to the question of how many stable 
models logic programs may have. We derive bounds on the maximum number 
of stable models in a program with $n$ atoms and use them to establish
lower and upper estimates on the performance of algorithms for computing 
all stable models.

Our main results concern propositional logic programs, called
$t$-programs, in which the number of literals in rules, including the
head, is bounded by a constant $t$. Despite their restricted syntax 
$t$-programs are of interest. Many logic programs that were proposed 
as encodings of problems in planning, model checking and combinatorics 
become propositional 2- or 3-programs after grounding. In general, 
programs obtained by grounding finite DATALOG$^\neg$ programs are 
$t$-programs, for some fixed, and usually small, $t$.

In the paper, for every $t\geq 2$, we construct an algorithm that 
computes all stable models of a $t$-program $P$ in time $O(m\alpha_
t^n)$, where $\alpha_t$ is a constant such that $\alpha_t < 2-1/2^t$. 
For 2-programs we obtain stronger results. We construct an algorithm 
that computes all stable models of a 2-program in time $O(m 3^{n/3})= 
O(m\times 1.44225^n)$. We note that $1.44225 < \alpha_2\approx1.61803$. 
Thus, this algorithm is indeed a significant improvement over the 
algorithm following from general considerations discussed above. 
We obtain similar results for a subclass of 2-programs consisting of 
programs that are purely negative and do not contain {\em dual} 
clauses. We also get significant improvements in the case when $t=3$.
Namely, we describe an 
algorithm that computes all stable models of a 3-program $P$ in 
time $O(m\times 1.70711^n)$. In contrast, since $\alpha_3\approx
1.83931$, the algorithm implied by the general considerations runs in 
time $O(m\times 1.83931^n)$. 

In the paper we also consider a general case where no bounds on the
length of a clause are imposed. We describe an algorithm to compute
all stable models of such programs. Its worst-case complexity is
slightly lower than that of the brute-force algorithm. 

It is well known that, by introducing new atoms, every logic program
$P$ can be transformed in polynomial time into a 3-program $P'$ that
is, essentially, equivalent to $P$: every stable model of 
$P$ is of the form $M'\cap \At$, for some stable model $M'$ of $P'$ and,
for every stable model $M'$ of $P'$, the set $M'\cap \At$ is a stable 
model of $P$. This observation might suggest that in order to design 
fast algorithms to compute stable models, it is enough to focus on the 
class of 3-programs. It is not the case. In the worst case, the number 
of new atoms that need to be introduced is of the order of the size of 
the original program $P$. Consequently, an algorithm to compute stable 
models that can be obtained by combining the reduction described above 
with an algorithm to compute stable models of 3-programs runs in time
$O(m 2^m)$ and is asymptotically slower than the brute-force 
approach outlined earlier. Thus, it is necessary to study algorithms for 
computing stable models designed explicitly for particular classes of
programs.

\section{Preliminaries}
\label{prel}

For a detailed account of logic programming and stable model semantics we 
refer the reader to \cite{gl88,ap90,mt93}. In the paper, we consider only 
the propositional case. For a logic program $P$, by $\At(P)$ we denote
the set of all atoms appearing in $P$. We define $\Lit(P) = \At(P) \cup 
\{\n(a)\colon a\in \At(P)\}$ and call elements of this set {\em literals}.
Literals $b$ and $\n(b)$, where $b$ is an atom, are {\em dual} to each
other. For a literal $\beta$, we denote its dual by $\n(\beta)$.

A {\em clause} is an expression $c$ of the form $p\lar B$ or $\lar B$, 
where $p$ is an atom and $B$ is a set of literals (no 
literals in $B$ are repeated). The clause of the first type is called 
{\em definite}. The clause of the second type is called a {\em 
constraint}. The atom $p$ is the {\em head} of $c$ and is denoted 
by $h(c)$. The set of atoms appearing in literals of $B$ is called the 
{\em body} of $c$. The set of all positive literals (atoms) in $B$ 
is the {\em positive body} of $c$, $b^+(c)$, in symbols. 
The set of atoms appearing in negated literals of $B$ is the {\em 
negative body} of $c$, $b^-(c)$, in symbols.

A {\em logic program} is a collection of clauses. If every clause of
$P$ is definite, $P$ is a {\em definite logic program}. If every clause
in $P$ has an empty positive body, that is, is {\em purely negative},
$P$ is a {\em purely negative} program. 
Finally, a logic program $P$
is a {\em $t$-program} if every clause in $P$ has no more than $t$
literals (counting the head).

A clause $c$ is a {\em tautology} if it is definite and $h(c)\in
b^+(c)$, or if $b^+(c)\cap b^-(c)\not=\emptyset$. A clause $c$ is a 
{\em virtual constraint} if it is definite and $h(c)\in b^-(c)$. We 
have the following result \cite{dix94b}.

\begin{proposition}
\label{smpl}
Let $P$ be a logic program and let $P'$ be the subprogram of $P$
obtained by removing from $P$ all tautologies, constraints and virtual
constraints. If $M$ is a stable model of $P$ then it is a stable model 
of $P'$.
\end{proposition}
 
Thanks to this proposition, when designing algorithms for computing
stable models we may restrict attention to definite programs without 
tautologies and virtual constraints. 

For a set of literals $L\subseteq \Lit(P)$, we define:
\[
L^+ =\{a\in \At(P)\colon a\in L\}\ \ \mbox{and}\ \ 
L^- =\{a\in \At(P)\colon \n(a)\in L\}.
\]
We also define $L^0 = L^+ \cup L^-$. A set of literals $L$ is {\em
consistent} if $L^+ \cap L^- =\emptyset$. A set of atoms $M 
\subseteq\At(P)$ is {\em consistent} with a set of literals $L\subseteq 
\Lit(P)$, if $L^+\subseteq M$ and $L^-\cap M =\emptyset$.

To characterize stable models of a program $P$ that are consistent 
with a set of literals $L\subseteq\Lit(P)$, we introduce a {\em
simplification} of $P$ with respect to $L$. By $[P]_L$ we denote the
program obtained by removing from $P$ 
\begin{enumerate}
\item every clause $c$ such that $b^+(c)\cap
L^-\not=\emptyset$ 
\item every clause $c$ such that $b^-(c)\cap
L^+\not=\emptyset$
\item every clause $c$ such that $h(c)\in L^0$
\item every occurrence of a literal in $L$ from the bodies of
the remaining clauses.
\end{enumerate}
The simplified program $[P]_L$ contains all information necessary to
reconstruct stable models of $P$ that are consistent with $L$.
The following result was obtained in \cite{dix94b} (we refer also to
\cite{vsnv95,ct99}).

\begin{proposition}
\label{key}
Let $P$ be a logic program and $L$ be a set of literals of $P$. 
If $M$ is a stable model of $P$ consistent with $L$, then $M\setminus 
L^+$ is a stable model of $[P]_L$.
\end{proposition}

Thus, to compute all stable models of $P$ that are consistent with 
$L$, one can first check if $L$ is consistent. If not, there are no stable
models consistent with $L$. Otherwise, one can compute all stable models 
of $[P]_L$, for each such model $M'$ check whether $M=M'\cup L^+$ is 
a stable model of $P$ and, if so, output $M$. This approach is the
basis of the algorithm to compute stable models that we present in the
following section.

\section{A high-level view of stable model computation}
\label{top-view}

We will now describe an algorithm $\stb(P,L)$ that, given a 
{\em definite} program $P$ and a set of literals $L$, outputs all 
stable models of $P$ that are consistent with $L$. 
The key concept we need is that of a complete collection. Let $P$
be a logic program. A nonempty collection $\mathcal A$ of nonempty
subsets of $\Lit(P)$ is {\em complete} for $P$ if every stable model
of $P$ is consistent with at least one set $A\in {\mathcal A}$.
Clearly, the collection ${\mathcal A}=\{\{a\}, \{\n(a)\}\}$, where
$a$ is an atom of $P$, is an example of a complete collection for
$P$. In the description given below, we assume that $\cmpl(P)$ is a
procedure that, for a program $P$, computes a collection of sets of 
literals that is complete for $P$. 

\vspace*{-0.2in}
\begin{tabbing}
\quad\=\quad\=\quad\=\quad\=\quad\=\quad\=\quad\=\quad\=\quad\=\quad\=
\quad\=\quad\=\\
$\stb(P,L)$\\
\medskip
(0)\>\>{\bf if} $L$ is consistent {\bf then}\\
(1)\>\>\>{\bf if} $[P]_L=\emptyset$ {\bf then}\\
(2)\>\>\>\>\>check whether $L^+$ is a stable model of $P$ and, if
so, output it\\
(3)\>\>\>{\bf else}\\
(4)\>\>\>\>${\mathcal A}:=\cmpl([P]_L)$;\\
(5)\>\>\>\>{\bf for every} $A\in {\mathcal A}$ {\bf do}\\
(6)\>\>\>\>\>$\stb(P,L\cup A)$\\
(7)\>\>\>{\bf end of} $\stb$.
\end{tabbing}

\begin{proposition}
\label{correct}
Let $P$ be a definite finite propositional logic program. For every $L
\subseteq \Lit(P)$, $\stb(P,L)$ returns all stable models of $P$ 
consistent with $L$.
\end{proposition}
Proof: We proceed by induction on $|\At([P]_L)|$. To start, let us 
consider a call to $\stb(P,L)$ in the case when $|\At([P]_L)|=0$ and
let $M$ be a set returned by $\stb(P,L)$. It follows that $L$ is
consistent and that $M$ is a stable model of $P$. Moreover, since 
$M=L^+$, $M$ is consistent with $L$. Conversely, let $M$ be a 
stable model of $P$ that is consistent with $L$. By Proposition 
\ref{key}, $M\setminus L^+$ is a stable model of $[P]_L$. Since $L$ is
consistent (as $M$ is consistent with $L$) and $[P]_L=\emptyset$, 
$M\setminus L^+ =\emptyset$. Since $M$ is consistent 
with $L$, $M=L^+$. Thus, $M$ is returned by $\stb(P,L)$. 

For the inductive step, let us consider a call to $\stb(P,L)$,
where $|\At([P]_L)| > 0$. Let $M$ be a set returned by this call. Then
$M$ is returned by a call to $\stb(P,L\cup A)$, for some $A\in 
{\mathcal A}$, where $\mathcal A$ is a complete family for $[P]_L$. 
Since elements of a complete family are nonempty and consist of
literals actually occurring in $[P]_L$, $|\At([P]_{L\cup A})| < 
|\At([P]_L)|$. By the induction hypothesis it follows that 
$M$ is a stable model of $P$ consistent with $L\cup A$ and, 
consequently, with $L$.

Let us now assume that $M$ is a stable model of $P$ consistent
with $L$. Then, by Proposition \ref{key}, $M\setminus
L^+$ is a stable model of $[P]_L$. Since ${\mathcal A}$ (computed in line 
(4)) is a complete collection for $[P]_L$, there is $A\in {\mathcal A}$ 
such that $M\setminus L^+$ is consistent with $A$. Since $A\cap L =
\emptyset$ (as $A\subseteq At([P]_L)$), $M$ is a stable model of $P$
consistent with $L\cup A$.
Since $|\At([P]_{L\cup A})| < 
|\At([P]_L)|$, by the induction hypothesis it follows that $M$ is output
during the recursive call to $\stb(P,L\cup A)$. \hfill$\Box$ 

We will now study the performance of the algorithm $\stb$. In our
discussion we follow the notation used to describe it. 
Let $P$ be a definite logic program and let $L\subseteq\Lit(P)$.
Let us consider the following recurrence relation:
\[
s(P,L) = \left\{
            \begin{array}{ll}
              1 & \mbox{if $[P]_L=\emptyset$ or $L$ is not consistent}\\
              \sum_{A\in{\mathcal A}} s(P,L\cup A)
                             \ \ \ & \mbox{otherwise.}
            \end{array}
         \right.
\]
As a corollary to Proposition \ref{correct} we obtain the following 
result.

\begin{corollary}
Let $P$ be a finite definite logic program and let $L\subseteq\Lit(P)$. 
Then, $P$ has at most $s(P,L)$ stable models consistent with $L$.
In particular, $P$ has at most $s(P,\emptyset)$ stable models.
\end{corollary}

We will use the function $s(P,L)$ to estimate not only the number of
stable models in definite logic programs but also the running time of 
the algorithm $\stb$. Indeed, let us observe that the total number of
times we make a call to the algorithm $\stb$ when executing
$\stb(P,L)$ (including the "top-level" call to $\stb(P,L)$) is
given by $s(P,L)$. We associate each execution of the instruction 
$(i)$, where $0\leq i\leq 5$,
with the call in which the instruction is executed. 
Consequently, each of these instructions is executed no more than 
$s(P,L)$ times during the execution of $\stb(P,L)$.

Let $m$ be the size of a program $P$. There are linear-time algorithms 
to check whether a set of atoms is a stable model of a program $P$.
Thus, we obtain the following result concerned with the performance of 
the algorithm $\stb$.

\begin{theorem}
If the procedure $\cmpl$ runs in time $O(t(m))$, where $m$ is the size
of an input program $P$, then executing the call $\stb(P,L)$, where
$L\subseteq \Lit(P)$, requires $O(s(P,L)(t(m)+m))$ steps in the worst case.
\end{theorem}

The specific bound depends on the procedure $\cmpl$, as it 
determines the recurrence for $s(P,L)$. It also depends on the 
implementation of the procedure $\cmpl$, as the implementation 
determines the second factor in the running-time formula derived 
above. 

Throughout the paper (except for Section \ref{gen}, where 
a different approach is used), we specify algorithms to compute 
stable models by describing particular versions of the procedure 
$\cmpl$. We obtain estimates on the running time of these 
algorithms by analyzing 
the recurrence for $s(P,L)$ implied by the procedure $\cmpl$. As a 
byproduct to these considerations, we obtain bounds on the 
maximum 
number of stable models of a logic program with $n$ atoms. 

\section{$t$-programs}
\label{tp}

In this section we will instantiate the general algorithm to compute 
stable models to the case of $t$-programs, for $t\geq 2$. To this
end, we will describe a procedure that, given a definite $t$-program
$P$, returns a complete collection for $P$.

Let $P$ be a definite $t$-program and let $x\lar \beta_1,\ldots,\beta_k$,
where $\beta_i$ are literals and $k\leq t-1$, be a clause in $P$. For 
every $i=1,\ldots,k$, let us define
\[
A_i=\{\n(x),\beta_1,\ldots,\beta_{i-1},\n(\beta_i)\}
\]
It is easy to see that the family ${\mathcal A} = \{\{x\}, A_1,\ldots, 
A_{k}\}$
is complete for $P$. We will assume that this complete collection is
computed and returned by the procedure $\cmpl$. Clearly, computing
$\cal A$ can be implemented to run in time $O(m)$. 

To analyze the resulting algorithm $\stb$, we use our general results
from the previous section. Let us define
\[
c_n= \left\{
            \begin{array}{ll}
              K_t & \mbox{\ \ if $0\leq n < t$}\\
              c_{n-1}+\ldots + c_{n-t} & \mbox{\ \ otherwise,}
            \end{array}
         \right.
\]
where $K_t$ is the maximum possible value of $s(P,L)$ for a $t$-program
$P$ and a set of literals $L\subseteq \Lit(P)$ such that $|\At(P)|-|L| 
\leq t$. We will prove that if $P$ is a $t$-program,
$L\subseteq \Lit(P)$, and $|\At(P)|-|L|\leq n$, then $s(P,L) \leq c_n$.
We proceed by induction on $n$. If $n < t$, then the assertion
follows by the definition of $K_t$. So, let us assume that $n\geq t$.
If $L$ is not consistent or $[P]_L=\emptyset$, $s(P,L) = 1\leq c_n$. 
Otherwise,
\[
s(P,L) = \sum_{A\in{\cal A}} s(P,L\cup A) \leq
c_{n-1}+ c_{n-2}+\ldots + c_{n-t}= c_n.
\]
The inequality follows by the induction hypothesis, the definition of
$\cal A$, and the monotonicity
of $c_n$. The last equality follows by the definition of $c_n$. Thus,
the induction step is complete.

The characteristic equation of the recurrence $c_n$ is $x^t=x^{t-1}+
\ldots+x+1$. Let $\alpha_t$ be the largest real root of this equation. 
One can show that for $t\geq 2$, $1 < \alpha_t < 2-1/2^t$. In particular, 
$\alpha_2\approx
1.61803$, $\alpha_3\approx 1.83931$, $\alpha_4\approx 1.92757$ and 
$\alpha_5\approx 1.96595$. The discussion in Section 
\ref{top-view} implies the following two theorems.

\begin{theorem}
\label{kp1}
Let $t$ be an integer, $t\geq 2$. There is an algorithm to compute
stable models of $t$-programs that runs in time $O(m \alpha_t^n)$, 
where $n$ is the number of atoms and $m$ is the size 
of the input program.
\end{theorem}

\begin{theorem}
\label{kp2}
Let $t$ be an integer, $t\geq 2$. There is a constant $C_t$ such that
every $t$-program $P$ has at most $C_t \alpha_t^n$ stable models,
where $n=|\At(P)|$.
\end{theorem}

Since for every $t$, $\alpha_t < 2$, we indeed obtain an improvement
over the straightforward approach. However, the scale of the
improvement diminishes as $t$ grows.

To establish lower bounds on the number of stable models and on the 
worst-case performance of algorithms to compute them, we define $P(n,t)$ 
to be a logic program such that $|\At(P)|=n$ and $P$ consists of all 
clauses of the form
\[
x\lar \n(b_1),\ldots,\n(b_{t}),
\]
where $x\in \At(P)$ and $\{b_1,\ldots,b_t\} \subseteq \At(P)\setminus
\{x\}$ are different atoms. It is easy to see that $P(n,t)$ is a
$(t+1)$-program with $n$ atoms and that stable models of $P(n,t)$ are 
precisely those subsets of $\At(P)$ that have $n-t$ elements. Thus, 
$P(n,t)$ has exactly ${n \choose t}$ stable models.

Clearly, the program $P(2t-1,t-1)$ is a $t$-program over the set of 
$2t-1$ atoms. Moreover, it has ${2t-1 \choose t}$ stable models. Let 
$kP(2t-1,t-1)$ be the logic program formed by the disjoint union of 
$k$ copies of $P(2t-1,t-1)$ (sets of atoms of different copies of
$P(2t-1,t-1)$ are disjoint). It is easy to see that $kP(2t-1,t-1)$ has 
${2t-1 \choose t}^k$ stable models. As an easy corollary of this 
observation we obtain the following result.

\begin{theorem}
\label{lb}
Let $t$ be an integer, $t\geq 2$. There is a constant $D_t$ such that
for every $n$ there is a $t$-program $P$ with at least $D_t\times{2t-1
\choose t}^{n/{2t-1}}$ stable models. 
\end{theorem}

This result implies that every algorithm for computing all stable
models of a $t$-program in the worst-case requires $\Omega({2t-1
\choose t}^{n/{2t-1}})$ steps, as there are programs for which at
least that many stable models need to be output. These lower bounds
specialize to approximately $\Omega(1.44224^n)$, $\Omega(1.58489^n)$, 
$\Omega(1.6618^n)$ and $\Omega(1.71149^n)$, for $t=2,3,4,5$, 
respectively.

\section{2-programs}
\label{2pr}

Stronger results 
can be
derived for more restricted classes of programs. 
We will now
study the case of 2-programs and prove the following two theorems.

\begin{theorem}
\label{2p1}
There is an algorithm to compute stable models of 2-programs that runs
in time $O(m 3^{n/3}) = O(m\times 1.44225^n)$, where $n$ is the 
number of atoms in $P$ and $m$ is the size of $P$.
\end{theorem}

\begin{theorem}
\label{2p2}
There is a constant $C$ such that every 2-program $P$ with $n$ atoms, 
has at most $C\times 3^{n/3}$ ($\approx C\times 1.44225^n$) stable models.
\end{theorem}

By Proposition \ref{smpl}, to prove these theorems it suffices to 
limit attention to the case of definite programs not containing
tautologies and virtual constraints. We will adopt this assumption 
and derive both theorems from general results presented in
Section \ref{top-view}. 

Let $P$ be a definite 2-program. We say that an atom $b\in 
\At(P)$ is a {\em neighbor} of an atom $a\in \At(P)$ if $P$ contains 
a clause containing both $a$ and $b$ (one of them as the head, the 
other one appearing positively or negatively in the body). By
$n(a)$ we will denote the number of neighbors of an atom $a$. Since we
assume that our programs contain neither tautologies nor virtual 
constraints, no atom $a$ is its own neighbor.

We will now describe the procedure $\cmpl$. The complete family
returned by the call to $\cmpl(P)$ depends on the program $P$. We list 
below several cases that cover all definite 2-programs without 
tautologies and virtual constraints. In each of these cases, we specify
a complete collection to be returned by the procedure $\cmpl$. 

\noindent
{\bf Case 1.} There is an atom, say $x$, such that $P$ contains a clause
with the head $x$ and with the empty body (in other words, $x$ is a fact 
of $P$). We define ${\cal A}=\{\{x\}\}$.
Clearly, every stable model of $P$ contains $x$. Thus, ${\cal A}$ 
is complete. 

\noindent
{\bf Case 2}. There is an atom, say $x$, that does not appear in the
head of any clause in $P$. We define
${\cal A} = \{\{\n(x)\}\}$.
It is well known that $x$ does not belong to {\em any} stable model of 
$P$. Thus, ${\cal A}$ is complete for $P$. 

\noindent
{\bf Case 3.} There are atoms $x$ and $y$, $x\not=y$, such that 
$x\lar y$ and at least one of $x \lar \n(y)$ and $y \lar \n(x)$ are 
in $P$. In this case, we set
${\cal A} = \{\{x\}\}$.
Let $M$ be a stable model of $P$. If $y\in M$, then $x\in M$ (due to
the fact that the clause $x\lar y$ is in $P$). Otherwise, $y\notin M$.
Since $M$ satisfies $x \lar \n(y)$ or $y \lar \n(x)$, it again follows 
that $x\in M$. Thus, ${\cal A}$ is complete.  

\noindent
{\bf Case 4.} There are atoms $x$ and $y$ such that $x\lar 
y$ and $y \lar x$ are both in $P$. We define
\[
{\cal A} = \{\{x,y\},\{\n(x),\n(y)\}\}.
\]
If $M$ is a stable model of $P$ then, clearly, $x\in M$ if and only
if $y\in M$. It follows that either $\{x,y\}\subseteq M$ or
$\{x,y\}\cap M=\emptyset$. Thus, ${\cal A}$ is complete for $P$.
Moreover, since $x\not=y$ ($P$ does not contain clauses of the form
$w \lar w$), each set in $\cal A$ has at least two elements.

\noindent
{\bf Case 5.} None of the Cases 1-4 holds and there is an atom, say 
$x$, with exactly one neighbor, $y$. Since $P$ does not contain 
clauses of the form $w\lar w$ and $w\lar\n(w)$, we have $x\not=y$. Moreover, 
$x$ must be the head of at least one clause (since we assume here
that Case 2 does not hold).\\
{\bf Subcase 5a.} $P$ contains the clause $x\lar y$. We define 
\[
{\cal A} = \{\{x,y\},\{\n(x),\n(y)\}\}.
\]
Let $M$ be a stable model of $P$. If $y\in M$ then, clearly, $x\in M$.
Since we assume that Case 3 does not hold, the clause $x\lar y$ is the 
only clause in $P$ with $x$ as the head. Thus, if $y\notin M$, then
we also have that $x\notin M$. Hence, $\cal A$ is complete.\\
{\bf Subcase 5b.} $P$ does not contain the clause $x\lar y$. 
We define
\[
{\cal A} = \{\{x,\n(y)\},\{\n(x),y\}\}.
\]
Let $M$ be a stable model of $P$. Since $x$ is the head of at least one 
clause in $P$, it follows that the clause $x\lar \n(y)$ belongs to $P$. 
Thus, if $y\notin M$ then $x\in M$. If $y\in M$ then, since $x\lar 
\n(y)$ is the only clause in $P$ with $x$ as the head, $x\notin M$.
Hence, $\cal A$ is complete. 

\noindent
{\bf Case 6.} None of the Cases 1-5 holds. Let $w\in \At(P)$ be an 
atom. By $x_1, \ldots,x_p$ we denote all atoms $x$ in $P$ such that 
$w\lar \n(x)$ or $x\lar\n(w)$ is a clause in $P$. Similarly, by 
$y_1,\ldots, y_q$ we denote all atoms $y$ in $P$ such that $y\lar w$ 
is a clause of $P$. Finally, by $z_1, \ldots,z_r$ we denote all atoms 
$z$ of $P$ such that $w\lar z$ is a clause of $P$. By our earlier 
discussion it follows that the sets $\{x_1,\ldots,x_p\}$, $\{y_1,
\ldots,y_q\}$ and $\{z_1,\ldots,z_r\}$, are pairwise disjoint and 
cover all neighbors of $w$. That is, the number of neighbors of $w$ 
is given by $p+q+r$. Since we exclude Case 5 here, $p+q+r\geq 2$. 
Further, since $w$ is the head of at least one edge (Case 2 does not
hold), it follows that $p+r\geq 1$

\noindent
{\bf Subcase 6a.} For some atom $w$, $q\geq 1$ or $p+q+r\geq 3$. Then,
we define
\[
{\cal A} = \{\{w,y_1,\ldots,y_q\},\{\n(w),
x_1,\ldots,x_p,\n(z_1),\ldots,\n(z_r)\}\}.
\]
It is easy to see that ${\cal A}$ is complete for $P$. Moreover, if
$q\geq 1$ then, since $p+r\geq 1$, each of the two sets in $\cal A$ 
has
at least two elements. If $p+q+r\geq 3$, then either each set in $\cal
A$ has at least two elements, or one of them has one element and the 
other one at least four elements.

\noindent
{\bf Subcase 6b.} Every atom $w$ has exactly two neighbors, and does
not appear in the body of any Horn clause of $P$. It follows that
all clauses in $P$ are purely negative. Let $w$ be an arbitrary atom 
in $P$. Let $u$ and $v$ be the two neighbors of $w$. The atoms $u$ and 
$v$ also have two neighbors each, one of them being $w$. Let $u'$ and 
$v'$ be the neighbors of $u$ and $v$, respectively, that are different 
from $w$. We define
\[
{\cal A} = \{\{\n(w),u,v\},\{\n(u),w,u'\},\{\n(v),w,v'\}\}.
\]
Let $M$ be a stable model of $P$. Let us assume that $w\notin M$. Since
$w$ and $u$ are neighbors, there is a clause in $P$ built of $w$ and
$u$. This clause is purely negative and it is satisfied by $M$. It 
follows that $u\in M$. A similar argument shows that $v\in M$, as well.
If $w\in M$ then, since $M$ is a stable model of $P$, there is a 2-clause 
$C$ in $P$ with the head $w$ and with the body satisfied by $M$. Since 
$P$ consists of purely negative clauses, and since $u$ and $v$ are the
only neighbors of $w$, $C= w\lar \n(u)$ or $C= w\lar \n(v)$. Let us 
assume the former. It is clear that $u\notin M$ (since $M$ satisfies the 
body of $C$). Let us recall that $u'$ is a neighbor of $u$. Consequently, 
$u$ and $u'$ form a purely negative clause of $P$. This clause is 
satisfied by $M$. Thus, $u'\in M$ and $M$ is consistent with
$\{\n(u),w,u'\}$. In the other case, when $C=w\lar \n(v)$, a similar
argument shows that $M$ is consistent with $\{\n(v),w,v'\}$. Thus,
every stable model of $P$ is consistent with one of the three sets in 
$\cal A$. In other words, ${\cal A}$ is complete. 

Clearly, given a 2-program $P$, deciding which of the cases described 
above holds for $P$ can be implemented to run in linear time. Once that 
is done, the output collection can be constructed and returned in linear 
time, too.

This specification of the procedure $\cmpl$ yields a particular
algorithm to compute stable models of definite 2-programs without 
tautologies and virtual constraints. To estimate its performance and 
obtain the bound on the number of stable models, we define 
\[
c_n= \left\{
            \begin{array}{ll}
              K & \mbox{if $0\leq n < 4$}\\
              \max\{c_{n-1}, 2c_{n-2}, c_{n-1}+c_{n-4}, 3c_{n-3}\} \ 
\ \ &
                            \mbox{otherwise,}
            \end{array}
         \right.
\]
where $K$ is the maximum possible value of $s(P,L)$, when $P$ is a 
definite finite propositional logic program, $L\subseteq \Lit(P)$ and 
$|\At(P)|-|L| \leq 3$. It is easy to see that $K$ is a constant that 
depends neither on $P$ nor on $L$. We will prove that $s(P,L) \leq c_n$,
where $n= |\At(P)|-|L|$. If $n\leq 3$, then the assertion follows by 
the definition of $K$. So, let us assume that $n\geq 4$. If $L$ is not
consistent or $[P]_L= \emptyset$, $s(P,L) = 1
\leq c_n$. Otherwise,
\[
s(P,L) = \sum_{A\in{\cal A}} s(P,L\cup A) \leq
\max\{c_{n-1}, 2c_{n-2}, c_{n-1}+c_{n-4}, 3c_{n-3}\} = c_n.
\]
The inequality follows by the induction hypothesis, the properties of
the complete families returned by $\cmpl$ (the cardinalities of sets
forming these complete families) and the monotonicity of $c_n$. 

Using well-known properties of linear recurrence relations, it is easy 
to see that $c_n = O(3^{n/3})= O(1.44225^n)$. Thus, Theorems \ref{2p1} 
and \ref{2p2} follow. 

As concerns bounds on the number of stable models of a 2-program, a
stronger (exact) result can be derived. Let 
\[
g_n = \left\{
            \begin{array}{ll}
              3^{n/3} & \mbox{\ \ if $n=0 \pmod{3}$}\\
              4\times 3^{(n-4)/3} & \mbox{\ \ if $n=1 \pmod{3}$, and $n>
1$}\\
              2\times 3^{(n-2)/3} & \mbox{\ \ if $n=2 \pmod{3}$}\\
              1 & \mbox{\ \ if $n=1$}
            \end{array}
         \right.
\]
Exploiting connections between stable models of purely negative 
definite 2-programs and maximal independent sets in graphs, and 
using some classic results from graph theory \cite{mm65} one can 
prove the following result.

\begin{corollary}
\label{max-stb}
Let $P$ be a 2-program with $n$ atoms. Then $P$ has no more than $g_n$
stable models.
\end{corollary}

The bound of Corollary \ref{max-stb} cannot be improved as there are
logic programs that achieve it. Let $P(p_1,\ldots,p_k)$, where for 
every $i$, $p_i\geq 2$, be a disjoint union of programs $P({p_1},1),
\ldots,P({p_k},1)$ (we discussed these programs in Section \ref{prel}).
Each program $P(p_i,1)$ has $p_i$ stable models. Thus, the number of 
stable models of $P(p_1,\ldots,p_k)$ 
is $p_1p_2\ldots p_k$. Let $P$ be a logic program with $n\geq 
2$ atoms and of the form $P(3,\ldots,3)$, $P(2,3,\ldots,3)$ or $P(4,3,
\ldots,3)$, depending on $n$(mod 3). It is easy to see that $P$ has 
$g_n$ stable models. In particular, it follows that our algorithm to
compute all stable models of 2-programs is must execute at least 
$\Omega(3^{n/3})$ steps in the worst case.

Narrowing the class of programs leads to still better bounds and 
faster algorithms. We will discuss one specific subclass of the class 
of 2-programs here. Namely, we will consider definite purely negative 
2-programs with no {\em dual} clauses (two clauses are called {\em dual}
if they are of the form $a\lar\n(b)$ and $b\lar \n(a)$). We denote the 
class of these programs by ${\cal P}_2^n$. Using the same approach as
in the case of arbitrary 2-programs, we can prove the following two
theorems.

\begin{theorem}
\label{2pn1}
There is an algorithm to compute stable models of 2-programs in the
class ${\cal P}_2^n$ that runs in time $O(m\times 1.23651^n)$, where $n$
is the number of atoms and $m$ is the size of an input program.
\end{theorem}

\begin{theorem}
\label{2pn2}
There is a constant $C$ such that every 2-program $P\in {\cal
P}_2^n$ has at most $C\times 1.23651^n$ stable models.
\end{theorem}

Theorem \ref{2pn2} gives an upper bound on the number of stable models
of a program in the class ${\cal P}_2^n$. To establish a lower bound,
we define $S_6$ to be a program over the set of atoms $a_1,\ldots,a_6$
and containing the rules (the arithmetic of indices is
performed modulo 6): $a_{i+1}\lar \n(a_i)$ and $a_{i+2}\lar \n(a_i)$,
$i=0,1,2,3,4,5$. The program $S_6$ has three stable models:
$\{a_0, a_1, a_3, a_4\}$, $\{a_1, a_2, a_4, a_5\}$ and $\{a_2, a_3, a_5, 
a_0\}$.

Let $P$ be the program consisting of $k$ copies of $S_6$, with
mutually disjoint sets of atoms. Clearly, $P$ has $3^k$ stable models.
Thus, there is a constant $D$ such that for every $n\geq 1$ there is a
program $P$ with $n$ atoms and with at least $D\times 3^{n/6}$
($\approx D\times 1.20094^n$) stable models.

\section{3-programs}
\label{2atoms}

We will now present our results for the class of 3-programs. 
Using similar techniques as those presented in the previous section, 
we prove the following two theorems.

\begin{theorem}
\label{3p1}
There is an algorithm to compute stable models of 3-programs that runs
in time $O(m\times 1.70711^n)$, where $m$ is the size of the input.
\end{theorem}

\begin{theorem}
\label{3p2}
There is a constant $C$ such that every 3-program $P$ has at most
$C\times 1.70711^n$ stable models.
\end{theorem}

The algorithm whose existence is claimed in Theorem \ref{3p1} is
obtained from the general template described in Section \ref{top-view}
by a proper instantiation of the procedure $\cmpl$ (in a similar way to
that presented in detail in the previous section for the case of
2-programs). 

The lower bound in this case follows from an observation made in 
Section \ref{tp} that there is a constant $D_3$ such that for every 
$n$ there is a 3-program $P$ such that $P$ has at least $D_3\times
1.58489^n)$ stable models (cf. Theorem \ref{lb}). Thus, every 
algorithm for computing all
stable models of 3-programs must take at least $\Omega(1.58489^n)$ 
steps in the worst case.

\section{The general case}
\label{gen}

In this section we present an algorithm that computes all stable models
of arbitrary propositional logic programs. It runs in time 
$O({m2^n}/{\sqrt{n}})$ and so, provides an improvement over the trivial 
bound $O(m2^n)$. However, our approach is quite different from that
used in the preceding sections. The key component of the algorithm is 
an auxiliary procedure $\stba(P,\pi)$. Let $P$ be a logic program and 
let $\At(P) =\{x_1,x_2,\ldots,x_n\}$. Given $P$ and a permutation $\pi$
of $\{1,2,\ldots,n\}$, the procedure $\stba(P,\pi)$ looks for
an index $j$, $1\leq j\leq n$, such that the set $\{x_{\pi(j)},\ldots,
x_{\pi(n)}\}$ is a stable model of $P$. Since no stable model of $P$ is 
a proper subset of another stable model of $P$, for any permutation 
$\pi$ there is at most one such index $j$. If such $j$ exists, the 
procedure outputs the set $\{x_{\pi(j)},
\ldots, x_{\pi(n)}\}$.

In the description of the algorithm $\stba$, we use the following 
notation.
For every atom $a$, by $\pos(a)$ we denote the list of all clauses 
which contain $a$ (as a non-negated atom) in their bodies, and by 
$\nn(a)$ a list of all clauses that contain $\n(a)$ in their bodies. 
Given a standard linked-list representation of logic programs, all
these lists can be computed in time linear in $m$.

Further, for each clause $C$, we introduce counters $p(C)$ and $n(C)$.
We initialize $p(C)$ to be the number of positive literals (atoms) in 
the body of $C$. Similarly, we initialize $n(C)$ to be the 
number of negative literals in the body of $C$. These counters are 
used to decide whether a clause belongs to the reduct of the program 
and whether it ``fires'' when computing the least model of the reduct. 

\vspace*{-0.2in}
\begin{tabbing}
\quad\=\quad\=\quad\=\quad\=\quad\=\quad\=\quad\=\quad\=\quad\=\quad\=
\quad\=\quad\=\\
$\stba(P,\pi)$\\
\medskip
(1)\>\>$M=\At(P)$;\\
(2)\>\>$Q:=$ set of clauses $C$ such that $p(C)=n(C)=0$;\\
(3)\>\>$\lm:=\emptyset$;\\
(4)\>\>{\bf for} $j=1$ {\bf to} $n$ {\bf do}\\
(5)\>\>\>\>{\bf while} $Q\not=\emptyset$ {\bf do}\\
(6)\>\>\>\>\>\>$C_0:=$ any clause in $Q$;\\
(7)\>\>\>\>\>\>mark $C_0$ as used and remove it from $Q$;\\
(8)\>\>\>\>\>\>{\bf if} $h(C_0)\notin \lm$ {\bf then}\\
(9)\>\>\>\>\>\>\>\>$\lm := \lm \cup \{h(C_0)\}$;\\
(10)\>\>\>\>\>\>\>\>{\bf for} $C\in pos(h(C_0))$ {\bf do}\\
(11)\>\>\>\>\>\>\>\>\>\>$p(C):= p(C)-1$;\\
(12)\>\>\>\>\>\>\>\>\>\>{\bf if} $p(C)= 0$ \& $n(C)=0$ \& $C$ not used
    {\bf then} add $C$ to $Q$;\\
(13)\>\>\>\>{\bf if} $\lm = M$ {\bf then} output $M$ and stop;\\
(14)\>\>\>\>$M:=M\setminus \{x_{\pi(j)}\}$;\\
(15)\>\>\>\>{\bf for} $C\in neg(x_{\pi(j)})$ {\bf do}\\
(16)\>\>\>\>\>\>$n(C):=n(C)-1$;\\
(17)\>\>\>\>\>\>{\bf if} $n(C)=0$ \& $p(C)=0$ \& $C$ not used {\bf 
then} 
           add $C$ to $Q$.
\end{tabbing}

Let us define $M_j=\{x_{\pi(j)},\ldots,x_{\pi(n)}\}$. Intuitively, 
the algorithm $\stba$ works as follows. In the iteration $j$ of the 
{\bf for} loop it computes the least model of the reduct $P^{M_j}$ 
(lines (5)-(12)). Then it tests whether $M_j = \lm(P^{M_j})$ (line (13)). 
If so, it outputs $M_j$ (it is a stable model of $P$) and terminates. 
Otherwise, it computes the reduct $P^{M_{j+1}}$. In fact the reduct 
is not explicitly computed. Rather, the number of negated literals 
in the body of each rule is updated to reflect the fact that we shift 
attention from the set $M_j$ to the set $M_{j+1}$ (lines (14)-(17)). 
The key to the algorithm is the fact that it computes reducts $P^{M_j}$ 
and least models $\lm(P^{M_j})$ in an incremental way and, so, tests $n$ 
candidates $M_j$ for stability in time $O(m)$ (where $m$ is the size 
of the program). 

\begin{proposition}
\label{gen1}
Let $P$ be a logic program and let $\At(P)=\{x_1,\ldots,x_n\}$. For
every permutation $\pi$ of $\{1,\ldots,n\}$, if $M=\{x_{\pi(j)},\ldots,
x_{\pi(n)}\}$ then the procedure $\stba(P,\pi)$ outputs $M$
if and only if $M$ is a stable model of $P$. Moreover,
the procedure $\stba$ runs in $O(m)$ steps, where $m$ is the size of 
$P$.
\end{proposition}

We will now describe how to use the procedure $\stba$ in an algorithm 
to compute stable models of a logic program. A collection ${\mathcal S}$ of 
permutations of $\{1,2,\ldots,n\}$ is {\em full} if every subset
$S$ of $\{1,2,\ldots,n\}$ is a final segment (suffix) of a
permutation in $\mathcal S$ or, more precisely, if for every subset 
$S$ of $\{1,2,\ldots,n\}$ there is a permutation $\pi\in {\mathcal S}$ 
such
that $S=\{\pi(n-|S|+1),\ldots,\pi(n)\}$. 

If $S_1$ and $S_2$ are of the same cardinality then they cannot occur 
as suffixes of the same permutation. Since there are
$n \choose {\lfloor n/2\rfloor}$ subsets of $\{1,2,\ldots,n\}$ of
cardinality $\lfloor n/2\rfloor$, every full family of
permutations must contain at least $n \choose{\lfloor n/2\rfloor}$
elements. An important property is that for every $n\geq 0$ there is a
full family of permutations of cardinality $n \choose{\lfloor
n/2\rfloor}$. An algorithm to compute such a minimal full set of
permutations, say ${\mathcal S}_{min}$, is described in \cite{kn73} (Vol.
3, pages 579 and 743-744). We refer to this algorithm as $\perm(n)$. 
The algorithm $\perm(n)$ enumerates all permutations in ${\mathcal 
S}_{min}$ by generating each next permutation entirely on 
the basis of the previous one. The algorithm $\perm(n)$ takes $O(n)$ 
steps to generate a permutation and each permutation is generated only 
once. 

We modify the algorithm $\perm(n)$ to obtain an algorithm to 
compute all stable models of a logic program $P$. Namely, each time 
a new permutation, say $\pi$, is generated, we make a call to 
$\stba(P,\pi)$. We call this algorithm $\stb^p$. Since ${n \choose
\lfloor n/2\rfloor} = \Theta(2^n/\sqrt{n})$ we have the following result.

\begin{proposition}
The algorithm $\stb^p$ is correct and runs in time
$O({m2^n}/{\sqrt{n}})$.
\end{proposition}
%

Since the program $P(n,\lfloor n/2\rfloor)$ has exactly ${n 
\choose{\lfloor n/2\rfloor}}$ stable models, every algorithm to compute 
all stable models of a logic program must take at least
$\Omega(2^n/\sqrt{n})$ steps.

\section{Discussion and conclusions}

We presented algorithms for computing stable models of logic 
programs with worst-case performance bounds asymptotically better than 
the trivial bound of $O(m 2^n)$. These are first results of that
type in the literature.
In the general case, we proposed an algorithm that runs in time 
$O(m2^n/\sqrt{n})$ improving the performance over the brute-force 
approach by the factor of $\sqrt{n}$. 
Most of our work, however, was concerned with algorithms for computing 
stable models of $t$-programs. We proposed an algorithm that computes 
stable models of $t$-programs in time $O(m\alpha_t^n)$, where $\alpha_t<
2-1/2^t$. We strengthened these results in the case of 2- and 3-programs. 
In the first case, we presented an algorithm that runs in time 
$O(m 3^{n/3})$ ($\approx O(m\times 1.44225^n)$). 
For the case of 3-programs, we 
presented an algorithm running in the worst case in time $O(m\times 
1.70711^n)$. 

In addition to these contributions, our work leads to several 
interesting questions. A foremost among them is whether our 
results can be further improved. First, we observe that in the case 
when the task is to compute {\em all} stable models, we already have proved
optimality (up to a polynomial factor) of the algorithms developed for 
the class of all programs and the class of all 2-programs. However, in 
all other cases there is still room for improvement --- our lower and 
upper bounds do not coincide.

The situation gets even more interesting when we want to compute
one stable model (if stable models exist) rather than all of them. 
Algorithms we presented here can, of course, be adapted to this case
(by terminating them as soon as the first model is found). Thus,
the upper bounds derived in this paper remain valid. But the lower
bounds, which we derive on the basis of the number of stable models input
programs may have, do not. In particular, it is no longer clear whether 
the algorithm we developed for the case of 2-programs remains optimal. 
One cannot exclude existence of pruning techniques that, in the case 
when the input program has stable models, would on occasion eliminate 
from considerations parts of the search space possibly containing some 
stable models, recognizing that the remaining portion of the search 
space still contains some. 

Such search space pruning techniques are possible in the case of 
satisfiability testing. For instance, the pure literal rule, sometimes 
used by implementations of the Davis-Putnam procedure, eliminates 
from considerations  parts of search space that may contain stable 
models \cite{ms85,kul99}. However, the part that remains is guaranteed
to contain a model as long as the input theory has one. No examples of 
analogous search space pruning methods are known in the case of stable 
model computation. We feel that nonmonotonicity of the stable model 
semantics is the reason for that but a formal account of this issue 
remains an open problem.

Finally, we note that many algorithms to compute stable models 
can be cast as instantiations of the general template introduced in 
Section \ref{top-view}. For instance, it is the case with the 
algorithm used in {\em smodels}. To view {\em smodels} in this way, 
we define the procedure complete as (1) picking (based on full 
lookahead) an atom $x$ on which the search will split; (2) computing 
the set of literals $A(x)$ by assuming that $x$ holds and by applying 
the unit propagation procedure of {\em smodels} (based, we recall on 
the ideas behind the well-founded semantics); (3) computing in the 
same way the set $A(\n(x))$ by assuming that $\n(x)$ holds; and (4) 
returning the family ${\cal A} =\{A(x), A(\n(x))\}$. This family is 
clearly complete. 

While different in some implementation details, the algorithm 
obtained from our general template by using this particular version 
of the procedure complete is essentially equivalent to that of {\em 
smodels}. By modifying our analysis in Section \ref{2pr}, one can show
that on 2-programs {\em smodels} runs in time $O(m\times 1.46558^n)$ 
and on purely negative programs without dual clauses in time $O(m
\times 1.32472^n)$. To the best of our knowledge these are first 
non-trivial estimates of the worst-case performance of {\em smodels}. 
These bounds are worse from those obtained from the algorithms we 
proposed here, as the techniques we developed were not designed with 
the analysis of {\em smodels} in mind. However, they demonstrate that 
the worst-case analysis of algorithms such as {\em smodels}, which is 
an important open problem, may be possible. 

\section*{Acknowledgments}

This material is based upon work supported by the National Science
Foundation under Grant No. 0097278.


\end{document}